\documentclass[lettersize,journal]{IEEEtran}
\usepackage{amsmath,amsfonts}
\usepackage{array}
\usepackage[caption=false,font=normalsize,labelfont=sf,textfont=sf]{subfig}
\usepackage{textcomp}
\usepackage{stfloats}
\usepackage{url}
\usepackage{verbatim}
\usepackage{graphicx}
\usepackage{cite}
\newcommand{\gnssfm}{GNSS-FM}

\begin{document}

\title{GNSS-FM: A Self-Supervised Foundation Model\\ for Daily GNSS Displacement Time Series}

\author{
Nick Teutschmann,
Laura Crocetti,
Fanny Lehmann,
Leonardo Trentini,
and Benedikt Soja%
\thanks{
N. Teutschmann, L. Crocetti, L. Trentini, and B. Soja are with the
Institute of Geodesy and Photogrammetry,
ETH Zurich, Zurich, Switzerland.
}%
\thanks{
F. Lehmann is with the ETH AI Center,
Zurich, Switzerland.
}
}

\maketitle

\begin{abstract}
Displacement time series from Global Navigation Satellite Systems (GNSS) are essential for a wide range of applications, including monitoring tectonic crustal deformations and investigating the different stages of the earthquake cycle. Machine learning methods have proven promising for GNSS applications; however, most remain fully supervised. This creates a bottleneck as labeled data are scarce, even though large amounts of unlabeled GNSS data are freely available. 
We present \gnssfm{}, a self-supervised foundation model for daily GNSS time series. The model uses a dual-stream input combining displacement and velocity-like increments, and is pretrained using a masked latent prediction objective with vector-quantized targets adapted from wav2vec 2.0, with several modifications for geodetic data. Pretrained on data from over 17,000 globally distributed GNSS stations, an analysis of the learned codebook suggests that the representations capture the main signal types in GNSS displacement data, including seismic offsets, tectonic drift, and seasonal patterns. The foundation model is later fine-tuned on two downstream tasks, namely 90-day displacement forecasting and seismic step localization, where it outperforms strong task-specific baselines in both cases. These results show that self-supervised pretraining is a promising approach for GNSS time series analysis.  
\end{abstract}

\begin{IEEEkeywords}
GNSS, foundation models, self-supervised learning, time series analysis,
geodesy, displacement monitoring, transformers, representation learning,
geophysical signals, Earth observation.
\end{IEEEkeywords}

\section{Introduction}
\label{sec:intro}

Global Navigation Satellite System (GNSS) networks provide continuous, high-precision measurements of Earth-surface displacement, making them a key tool for monitoring crustal deformation. GNSS displacement time series capture a wide range of patterns, from long-term tectonic motion and seasonal loading to abrupt seismic offsets caused by earthquakes \cite{Freymueller2017}. As GNSS networks have grown globally, datasets now contain displacements from tens of thousands of stations spanning multiple decades \cite{blewitt2018gps_explosion}. This scale makes data-driven approaches attractive because they can learn from a rich global dataset that covers diverse tectonic, climatic, and station-quality conditions. 

Despite this availability of large amounts of unlabeled data, i.e., data without explicit labels, most existing machine learning methods for GNSS time series are task-specific: they are trained from scratch for a single task, such as discontinuity detection or GNSS displacement forecasting \cite{crocetti2021discontinuity,chen_improved_2023}. While forecasting does not require additional labels, many other geophysically relevant tasks do. These labels are often limited or incomplete: discontinuity detection can depend on incomplete external catalogs, and slow slip event catalogs are sparse and usually restricted to the largest events \cite{crocetti2021discontinuity,costantino2023sse}. Instead, the GNSS data needed to pretrain a more general model is freely available and continuously growing. This gap between data availability and label shortage is the main motivation for this work. 

Self-supervised learning is a natural way to exploit large amounts of unlabeled data. In speech and vision, large encoders pretrained on unlabeled data using masked prediction objectives have been shown to learn representations that transfer well to many downstream tasks \cite{devlin2019bert, he2022mae}. The wav2vec 2.0 model \cite{baevski_wav2vec_2020} showed that this approach is especially effective when prediction targets are formed by discretizing the latent representation with a vector quantizer. This prevents the model from simply fitting noise and instead encourages it to capture recurring structure. SeisLM \cite{liu_seislm_2024} showed that the same approach transfers to geophysical data, producing a foundation model for seismic time series that improves performance across several downstream tasks. 

However, directly applying these architectures to GNSS is not straightforward. Unlike seismic waveforms or speech, GNSS displacement series are non-stationary over multi-year windows, and combine slow tectonic drift, seasonal loading, abrupt step offsets, and colored measurement noise in a single signal. Data gaps are also common. At the same time, the displacement history within a window contains information that is important for downstream tasks, so it should not simply be removed by differencing or detrending. This creates a trade-off for self-supervised pretraining: the model should learn local changes, while still maintaining longer-range deformation context. Existing architectures can therefore not be applied unchanged, but need to be adapted to the structure of GNSS displacement time series. 

We present the \emph{GNSS Foundation Model} (\gnssfm{}), a self-supervised foundation model for daily GNSS displacement time series that adapts the wav2vec 2.0 framework to geodetic data, pretrained on data from over 17,000 globally distributed stations provided by the Nevada Geodetic Laboratory \cite{blewitt2018gps_explosion}. 

Our main contributions are:
\begin{itemize}
    \item a dual-stream input representation combining displacement and velocity-like increments, giving the model access to both long-range deformation context and local dynamics,
    \item a masked latent prediction pretraining framework with reliability-aware masking and loss weighting, conditioning on station metadata, and a collapse-resistant grouped spherical quantizer.
\end{itemize} 
We also provide a qualitative analysis suggesting that the learned representations capture the main signal types present in GNSS data, including seismic offsets, tectonic drift, and seasonal signals. We evaluate the pretrained model on 90-day displacement forecasting and seismic step localization, where it outperforms strong task-specific baselines in both cases. 

\section{Related Work}
\label{sec:related}

\subsection{Self-Supervised Learning for Sequential Signals}

The pretrain-finetune approach has become standard in natural language processing and computer vision, where large encoders trained on unlabeled data with masked prediction objectives learn representations that transfer well to many downstream tasks \cite{devlin2019bert, he2022mae, bao2022beit}. For continuous signals, directly reconstructing masked values tends to make the model focus on small variations (i.e., noise and high-frequency details) rather than the underlying structure. This happens because small numerical differences tend to dominate the error signal. A common solution is to predict simpler discrete targets instead of the original continuous values, reducing sensitivity to noise and encouraging the model to learn more meaningful patterns, an observation that has also been made for noisy SAR imagery in remote sensing \cite{wang2024fgmae}. The wav2vec 2.0 model \cite{baevski_wav2vec_2020} is based on that principle and uses a convolutional encoder that maps the input to latent features. These latent features are then quantized to produce discrete targets. Here, the \emph{codebook} is a learned dictionary of representative latent patterns, and quantization assigns each continuous latent feature to one entry in this dictionary. Some of the latent features are masked, and a transformer builds representations from the visible context. The model is then trained to predict the correct quantized target for the masked part of the signal. A diversity loss encourages the model to use the full codebook and avoid collapse. HuBERT \cite{hsu2021hubert} replaces the quantizer with cluster assignments that are updated during training, which improves stability. WavLM \cite{chen2022wavlm} adds denoising objectives and further architectural improvements. 

The key idea that transfers to GNSS is that predicting discrete targets encourages the encoder to capture recurring patterns rather than fitting noise. This is especially useful for geodetic data, where much of the daily variability is measurement noise rather than a real deformation signal. 

\subsection{Self-Supervised and Foundation Models for Time Series}

Self-supervised methods for time series have grown rapidly, though most research focuses on high-frequency sensor data or financial and meteorological series, which are quite different from geodetic data. Contrastive methods teach models to recognize that different augmented versions of the same time window should be represented in a similar way \cite{yue2022ts2vec}. Masked modeling methods apply the same masked prediction idea to time series, either by reconstructing raw values \cite{zerveas2021transformer} or predicting discrete targets \cite{gui_vector_2024, labach_effective_2022}. In Earth observation, similar masked reconstruction objectives have been applied to optical satellite image time series \cite{dumeur2024ubarn}. VQ-MTM \cite{gui_vector_2024} is similar in design to our approach: it applies a random-projection quantizer to electroencephalogram time series and shows that discrete targets help for noisy, non-stationary signals beyond speech. 

Several recent models apply pretraining specifically to time series forecasting. Chronos \cite{ansari2024chronos} discretizes time series values into a vocabulary and trains transformer language models with cross-entropy, achieving strong zero-shot forecasting across many datasets. TimesFM \cite{das2023timesfm} and TimeGPT \cite{garza2023timegpt} follow a similar idea with decoder-only architectures. These models share the general goal of learning from large amounts of diverse data, but they are designed mainly for forecasting and do not address the specific properties of geodetic time series. 

Recent foundation models for Earth-system and geospatial data are closer to geodetic applications than general time series foundation models. Aurora introduces a foundation model pretrained on heterogeneous Earth-system data and finetuned for several forecasting tasks, including weather, air quality, ocean waves, and tropical cyclone tracks \cite{bodnar2025foundation}. Building on this, Trentini et al. \cite{trentini2026gnss} explore integrating GNSS zenith wet delays into the same framework to improve moisture-related forecasting. The Earth System Foundation Model builds on this by improving the flexibility of the input representation and training protocols. In particular, it uses variable-wise tokenization and structured masking strategies to handle partially observed variables, sparse satellite observations, and station data within one backbone, while preserving dependencies between physical variables \cite{ozdemir2026esfm}. AlphaEarth Foundations follows a related direction in Earth observation by learning geospatial embeddings from multi-source spatial, temporal, and measurement context for mapping from sparse labels \cite{brown2025alphaearth}. 

\begin{figure*}[!t]
    \centering
    \includegraphics[width=\linewidth]{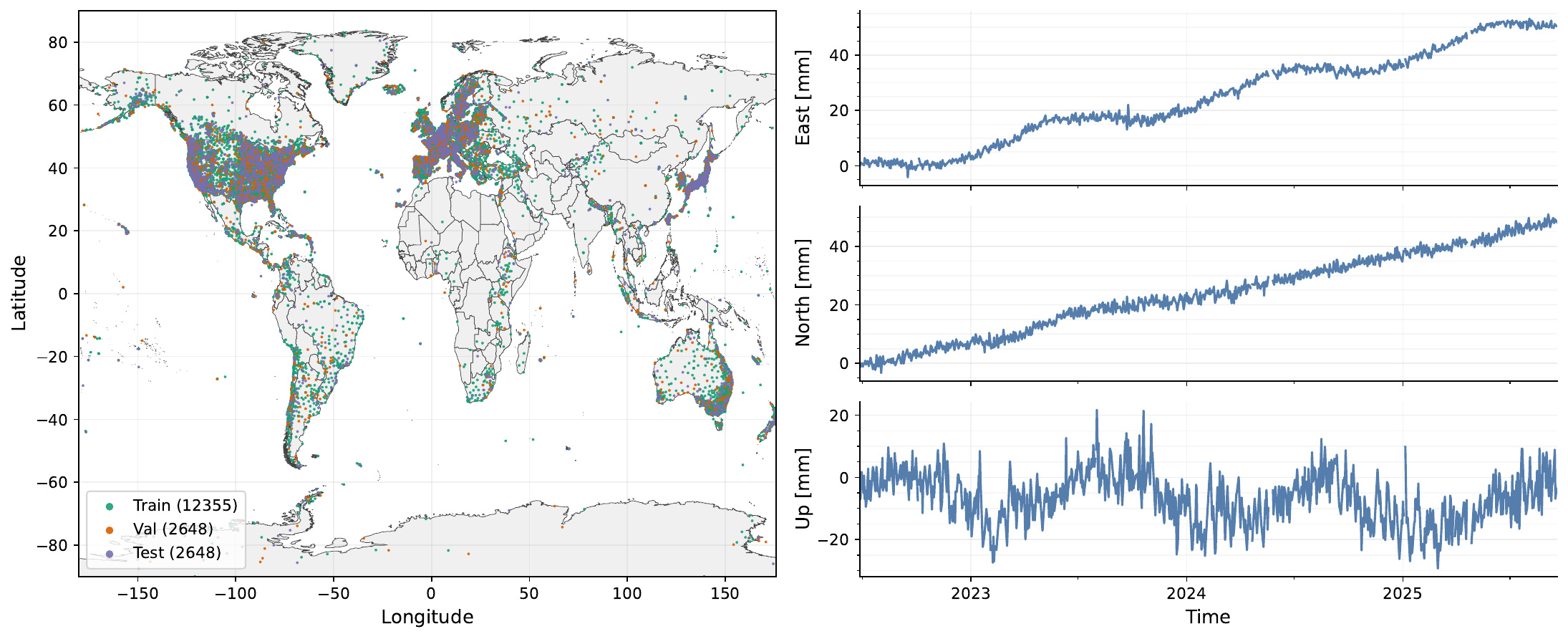}
    \caption{Dataset overview. Left: Global distribution of GNSS stations in the geographically stratified training, validation, and test splits. Right: Example of a daily GNSS displacement time series (east, north, and up component) for station OOSE.}
    \label{fig:dataset_overview}
\end{figure*}

\subsection{Foundation Models for Seismic Waveforms}

SeisLM \cite{liu_seislm_2024} is, to our knowledge, the most directly related prior work. It adapts wav2vec 2.0 to seismic waveforms, is pretrained on multiple seismic datasets and shows strong transfer capability for phase picking and event classification. Our work follows the same general approach but applies it to a single dataset of daily GNSS displacement time series, which differ from seismic waveforms in sampling rate, signal structure, and downstream task requirements. Parts of our implementation are adapted from SeisLM. The main adjustments needed for GNSS, namely the dual-stream representation, reliability-aware conditioning, and a quantizer designed for lower-frequency non-stationary data, are the central contribution of this work. 

Machine learning and deep learning methods have been applied to GNSS for discontinuity detection \cite{crocetti2021discontinuity}, denoising \cite{mastella_denoising_2025, bachelot2023gnssgnn}, environmental-loading-related residual or displacement correction \cite{ruttner2022modeling,crocetti2025correction}, and displacement forecasting \cite{simsek_modelling_2024, kiani_shahvandi_inclusion_2022}. All of these are trained for a single specific task. Our approach is different because the same backbone, pretrained without task-specific labels, is finetuned to multiple downstream tasks. 

\subsection{GNSS Displacement Time Series Modeling}

GNSS displacement series are typically modeled as a combination of a trend, annual and semi-annual periodic terms, and a colored noise process \cite{williams_effect_2003, dong_seasonal_2002}. In geodetic reference-frame realizations such as ITRF2020, station motion is modeled more explicitly through piecewise linear trajectories with discontinuities, seasonal terms, and post-seismic deformation for sites affected by major earthquakes \cite{Altamimi2023ITRF2020}. 

Standard preprocessing steps include outlier removal, step correction, and spatial and temporal filtering to reduce shared noise \cite{wdowinski_southern_1997, dong2006kle}. For forecasting, hybrid models that combine variational mode decomposition with recurrent or transformer networks have shown improvements over simpler baselines \cite{chen_improved_2023, jiao_noise-resilient_2025}. For step detection, Crocetti et al. \cite{crocetti2021discontinuity} showed that random forest classifiers work well on short 21-day windows. 

\section{Data and Preprocessing}
\label{sec:data}

\subsection{Dataset}

We use daily GNSS displacement time series from the Nevada Geodetic Laboratory \cite{blewitt2018gps_explosion}, which provides east, north, and up displacement components. Starting from 23,042 stations, we apply quality filters that remove stations with fewer than one year of data, an observed fraction below 70\%, or an internal gap longer than three years. This leaves 17,652 stations with a median span of 4,048 days and a median coverage fraction of 0.949, totaling 73.4 million observed station-days. The stations are divided into training (70\%), validation (15\%), and test (15\%) sets using a geographically stratified split, so that each subset has a broad global coverage. Figure~\ref{fig:dataset_overview} illustrates the resulting spatial distribution of stations across the three splits and shows an example GNSS displacement time series for one station. 

\subsection{Preprocessing}
\label{subsec:preprocessing}

We summarize the main preprocessing steps here and present more details in Appendix \ref{app:preprocessing}. Outliers are detected separately for each component with an iterative procedure that combines a Hampel filter, large nearest-neighbor differences, and checks of the displacement before and after the suspected outlier, so that isolated spikes are removed without removing persistent step changes \cite{pearson2016generalized,davies1993identification}. Detected outlier samples are masked and treated as gaps. The cleaned displacement series is then converted to daily increments by first differencing. Equipment-related offsets from the Nevada Geodetic Laboratory event catalog \cite{blewitt2018gps_explosion} are handled in this increment space rather than by shifting the full displacement time series. Since a displacement offset appears as a single spike after first differencing, the affected daily increment is masked and then filled during gap filling. The same catalog is also used to derive the targets for seismic step localization, restricted to seismic events with a local detectability score $Z>3$, where $Z$ is the norm of the largest three-component daily increment within a $\pm 3$-day window around the catalog date, normalized by the robust variability of a preceding 30-day reference window. Equipment-related offsets are excluded and not used as positive targets. 

Gaps in the daily-increment series are filled using two methods. Short gaps of up to three days are filled by linear interpolation, while longer gaps use a bootstrap-based method that samples chunks from nearby context to maintain realistic noise characteristics \cite{kunsch1989jackknife,lall1996nearest}. For a gap of length $L$, this context is taken from valid samples within at least 200 days on either side of the gap and expanded to three times the gap length for longer gaps. This provides enough local samples for stable resampling while keeping the synthesized values tied to the surrounding time series behavior. The displacement stream is then reconstructed from the filled increments. Each day is assigned a reliability label according to how the value was obtained: original observation, short-gap linear interpolation, bootstrap synthesis, or padding. These labels are mapped to weights of 1.0, 0.6, 0.2, and 0.0, respectively. They reflect decreasing confidence in the values and are used in two ways: to weight the loss during training, and as a conditioning signal fed into the model to introduce a measure of data quality. 

\subsection{Dual-Stream Representation}
\label{sec:dual_stream}

The model receives two different views of the same displacement series: (1) the displacement stream, and (2) the velocity stream. The velocity stream is formed by first-order differences of the displacement series. These values are therefore derived daily displacement increments, not directly observed velocity measurements, but are referred to as the velocity stream in the following for the sake of clarity. This stream reduces non-stationarity and highlights local changes. The displacement stream maintains the absolute displacement level and long-range context, including tectonic drift and seasonal patterns. Both streams are normalized independently using the median and the scaled median absolute deviation. For each component series $y_t$ in each stream, we compute $m=\mathrm{median}(y)$, $s=1.4826\,\mathrm{median}(|y_t-m|)$, and $z_t=(y_t-m)/s$. The factor $1.4826$ makes the median absolute deviation consistent with the standard deviation under normally distributed noise. An inverse hyperbolic sine transformation, $\tilde{z}_t=\mathrm{asinh}(z_t)$, is then applied to compress large values without clipping them.

\subsection{Windowing}

All models use fixed-length windows of $W = 512$ days as input. For daily GNSS displacement time series, this window length covers more than one annual cycle, allowing the model to observe seasonal loading signals together with longer-term trends and abrupt offsets within the same sample. Each window must contain at least 80\% originally observed days to limit training on interpolated data. For pretraining and forecasting, windows are non-overlapping to prevent data leakage. For detection, windows may overlap by up to 20\% so that events near the boundaries of the window are still captured with sufficient context on both sides. 

For seismic step localization, training windows that contain events, i.e., windows with positive labels, are augmented by random time shifts of up to 12 days. Detection training also uses balanced batches with one positive-label window per batch so that events are seen regularly despite the strong class imbalance. 

\begin{figure*}[!t]
  \centering\includegraphics[width=\linewidth]{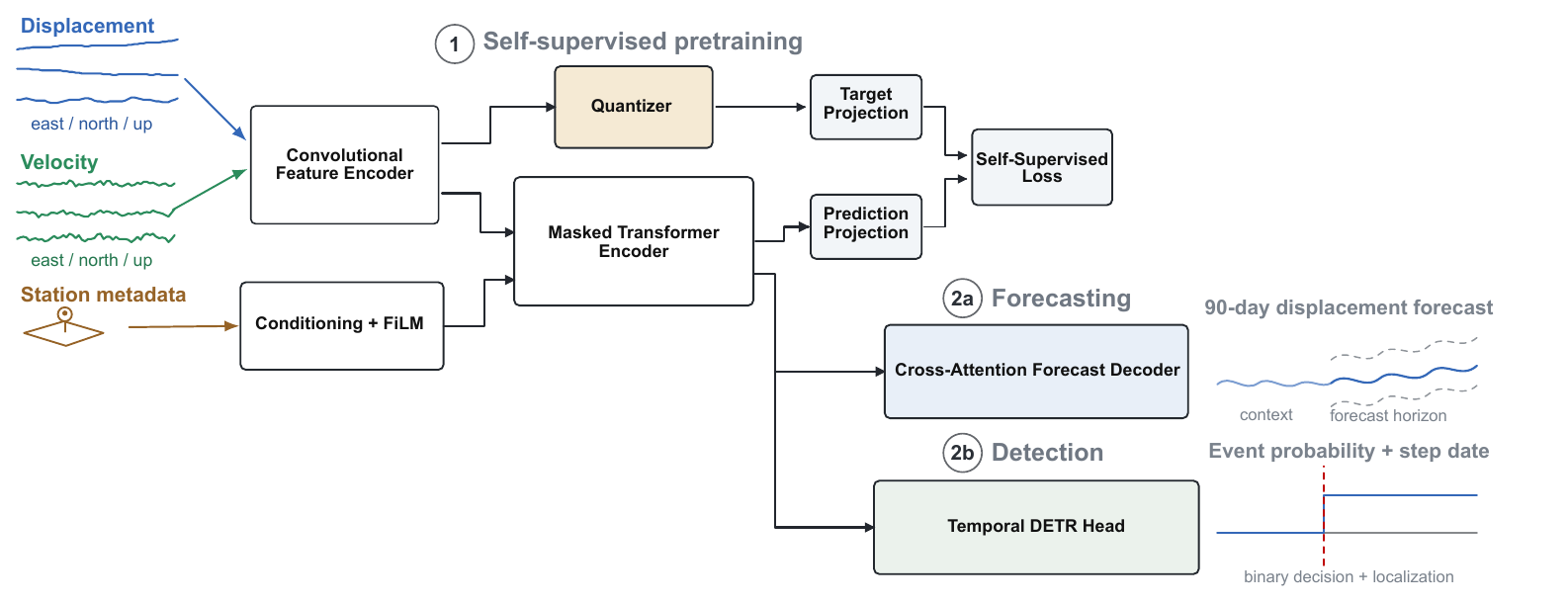}
  \caption{Overview of \gnssfm{}. The model takes displacement and velocity time series as inputs, together with station metadata. During self-supervised pretraining \textcircled{\tiny{1}}, a shared convolutional feature encoder and masked transformer encoder are trained to predict quantized representations of masked targets using a contrastive objective. After pretraining, the same encoder is reused for two downstream tasks: \textcircled{\tiny{2a}} 90-day displacement forecasting with a cross-attention decoder, and \textcircled{\tiny{2b}} seismic step localization with a temporal DETR head. Station metadata modulates the transformer through FiLM, while remaining separate from the quantizer targets.}
  \label{fig:architecture}
\end{figure*}

\section{Method}
\label{sec:method}

\subsection{\gnssfm{} Structure}

\gnssfm{} adapts the masked latent prediction of wav2vec 2.0 \cite{baevski_wav2vec_2020} and the structural framework of SeisLM \cite{liu_seislm_2024} and redesigns them for the specific characteristics of daily geodetic time series. Given windowed GNSS time series, the model extracts convolutional features, masks a subset of feature-time positions, encodes the visible context with a transformer, and predicts vector-quantized targets at the masked positions using a contrastive loss. After pretraining, the convolutional encoders and the transformer encoder are reused for downstream tasks, while the quantizer and contrastive projection heads are discarded. Forecasting is performed with a cross-attention decoder, and seismic step localization with a temporal Detection Transformer (DETR) head \cite{Carion2020EndtoEndOD}. Figure \ref{fig:architecture} gives an overview of the full model and its adaptation to the two downstream tasks. The full dual-stream pretraining model contains 359 million parameters in total. 

\subsection{Convolutional Feature Encoder}

Each input stream is processed by a separate strided 1D convolutional encoder with five layers, kernel sizes $(5,3,3,3,3)$, and strides $(2,2,1,1,1)$. All layers use 256 channels, layer normalization \cite{ba2016layernorm}, and Gaussian Error Linear Unit activations \cite{hendrycks2016gelu}. For a 512-day input window, this produces a feature sequence of length $T' = 120$, corresponding to roughly 4 days per feature step and a receptive field of 33 days. The receptive field is computed recursively as
\begin{equation}
    R_l = R_{l-1} + (k_l - 1)J_{l-1}, \qquad
    J_l = J_{l-1}s_l,
\end{equation}
where $R_l$ is the receptive field after layer $l$, $J_l$ is the stride relative to the input, and $k_l$ and $s_l$ are the kernel size and stride of layer $l$, with $R_0 = J_0 = 1$. The downsampling suppresses high-frequency noise while maintaining enough temporal resolution for the downstream tasks. The two encoders have separate parameters so they can specialize for their respective input streams. 

\subsection{Dual-Stream Transformer Encoder}

After convolution, each stream is layer-normalized and projected to the hidden dimension $H~=~768$ of the transformer. Positional information is added using wav2vec 2.0-style convolutional positional embeddings \cite{baevski_wav2vec_2020}, applied separately to each stream. Both streams are then processed by parallel self-attention stacks with the same depth ($L = 12$ layers, $n_h = 12$ heads, feed-forward dimension $4H$). Bidirectional cross-attention blocks are inserted after every second self-attention layer so that the two streams can exchange information. Residual updates are scaled by 0.8 and per-channel LayerScale parameters \cite{Touvron_2021_ICCV} are initialized close to zero. Both choices helped stabilize training in the early epochs. 

\subsection{Conditioning on Metadata and Data Quality}

We condition the transformer on station metadata and reliability information. Station metadata, consisting of latitude, longitude, and ellipsoidal height encoded as sine-cosine pairs, are projected into a conditioning vector and used to modulate every transformer layer through FiLM \cite{perez2018film}. The reliability label of each feature-time step is processed in the same way, so the encoder knows whether each value is an original observation, interpolated, bootstrap-synthesized, or padded (see Section \ref{subsec:preprocessing}), with corresponding reliability weights of 1.0, 0.6, 0.2, and 0.0. A learnable global gate controls the overall conditioning strength and is gradually increased during the first quarter of training to keep early optimization stable. 

\subsection{Vector Quantization}

Prediction targets are produced by discretizing the convolutional features before masking and conditioning. Using discretized targets derived from raw unconditioned features prevents the model from taking shortcuts. The quantizer has $G = 4$ independent codebook groups with $V = 320$ learnable codevectors each. The first two groups receive velocity features, and the last two receive displacement features. Before quantization, each group has its own learnable linear projection from the convolutional feature space into its codebook subspace. Each projected feature is then compared with the 320 learnable codevectors in its group, and one codevector is selected as the target representation. All vectors are $\ell_2$-normalized before computing cosine-similarity logits, which makes assignments independent of magnitude and prevents large seismic steps from dominating the surrounding signal. 

During training, group assignments use Gumbel-Softmax \cite{jang2017gumbel} with a temperature that decreases from 1.5 to 0.7, while the logit scale increases from 1.5 to 3.0. This encourages the model to explore many codes early in training and select sharper assignments later. Codevectors that are rarely chosen are periodically replaced with normalized features from the current batch to prevent collapse. The codebook is initialized using spherical $k$-means on a small subset of the training data. 

\subsection{Pretraining Objective}

Short continuous sub-windows of the feature sequence are masked before being passed to the transformer. A masking curriculum increases the masking probability from 12\% to 50\% and the span length from 8 to 12 over the first 30 epochs. An additional tail-masking component masks the last few positions of a window, which trains the model to extrapolate rather than only fill in gaps. Both streams are always masked at the same positions, so the model cannot recover masked velocity from the visible displacements or masked displacement from the visible velocities. 

The main training loss is an Information Noise-Contrastive Estimation (InfoNCE) contrastive objective \cite{oord2018cpc} at masked positions: 
\begin{equation}
\mathcal{L}_{\mathrm{ctr}} = -\frac{1}{|\mathcal{M}|}
\sum_{t \in \mathcal{M}} \log \frac{ e^{\,\mathrm{sim}(c_t, q_t)\,/\,\tau} }{ e^{\,\mathrm{sim}(c_t, q_t)\,/\,\tau} + \displaystyle\sum_{j=1}^{K} e^{\,\mathrm{sim}(c_t, q_{t,j}^{-})\,/\,\tau} },
  \label{eq:infonce}
\end{equation}
where $\mathcal{M}$ is the set of masked positions, $c_t$ is the projected transformer output, $q_t$ is the quantized target, $\{q_{t,j}^{-}\}$ are $K = 50$ negatives sampled from other positions and other batches, $\mathrm{sim}$ is the cosine similarity, and $\tau = 0.1$ is a fixed temperature. Here, negatives are quantized targets taken from other feature steps, so the model must distinguish the correct target at step $t$ from incorrect targets selected from elsewhere. To increase the diversity of negatives without increasing batch size, we use a memory queue of past targets. This gives a large pool of negatives while keeping the implementation fixed across all experiments. Each loss term is weighted by the reliability label of the corresponding position, so synthesized or interpolated values contribute less than original observations. The full pretraining objective is 
\begin{equation}
\begin{aligned}
\mathcal{L} ={}&
\mathcal{L}_{\mathrm{ctr}}
+ \lambda_{\mathrm{div}} \mathcal{L}_{\mathrm{div}}
+ \lambda_{\mathrm{aux}}(s)\, \mathcal{L}_{\mathrm{aux}}
+ \lambda_{\mathrm{orth}} \mathcal{L}_{\mathrm{orth}}
+ \lambda_{\mathrm{tv}} \mathcal{L}_{\mathrm{tv}} \\
&+ \lambda_{\mathrm{vh}}(s)\, \mathcal{L}_{\mathrm{vh}}
+ \lambda_{\mathrm{vkl}}(s)\, \mathcal{L}_{\mathrm{vkl}}
+ \lambda_{\mathrm{lat}}(s)\, \mathcal{L}_{\mathrm{lat}}
+ \lambda_{\mathrm{sep}}\, \mathcal{L}_{\mathrm{sep}} \\
&- \lambda_{\mathrm{usage}} \mathcal{H}_{\mathrm{usage}}
- \lambda_{\mathrm{tok}} \mathcal{H}_{\mathrm{tok}} \, 
\end{aligned}
\label{eq:full_pretrain_loss}
\end{equation}
where $\mathcal{L}_{\mathrm{ctr}}$ is the main contrastive loss, $\mathcal{L}_{\mathrm{div}}$ encourages broad codebook usage \cite{baevski_wav2vec_2020}, $\mathcal{L}_{\mathrm{aux}}$ is a small auxiliary reconstruction term, and the remaining terms regularize code usage and latent structure to reduce collapse. The coefficients that depend on $s$ are scheduled during training, where $s$ refers to the training step. Key hyperparameters are listed in Table \ref{tab:hyperparams}, and full definitions of the individual terms are given in Appendix \ref{app:pretraining_details}. 

\begin{table}[!t]
  \centering
  \caption{Key hyperparameters of the \gnssfm{} base model.}
  \label{tab:hyperparams}
  \begin{tabular}{|l|l|}
    \hline
    Component   & Setting \\
    \hline
    Window      & $T=512$ days,\; $T'=120$ features \\
    Transformer & $L=12$,\; $H=768$,\; $n_h=12$ heads \\
    Masking     & $p=0.50$, span 12;\; tail $p=0.08$, span 8 \\
    Quantizer   & $G=4$ groups,\; $V=320$ codes/group \\
    Negatives   & $K=50$,\; memory queue 16,384 \\
    Optimizer   & AdamW,\; learning rate $10^{-4}$  \\
    Training    & 40 epochs,\; 7 GPUs,\; batch 448 \\
    \hline
  \end{tabular}
\end{table}

\subsection{Downstream Task Adaptation} 

For displacement forecasting, each 512-day window is split into a 422-day context and a 90-day forecast horizon. The pretrained model encodes the context window and reconstructs the masked forecast horizon, and a task-specific cross-attention decoder combines these representations to predict the future displacement trajectory. The cross-attention decoder has 4 layers, 8 attention heads, model dimension 256, feed-forward expansion factor 4, and dropout 0.1. The decoder uses one query per forecast day and per displacement component, so each future time step receives its own representation. Each decoder layer performs cross-attention over three inputs: the encoded context, the reconstructed horizon of the pretrained transformer, and the preprocessed displacement values. Finetuning is achieved with Low-Rank Adaptation \cite{hu2021lora} applied to the main transformer linear layers, and the last four transformer layers are additionally unfrozen. With this adaptation strategy, the forecasting model contains 382 million parameters, of which 70.4 million are trainable during finetuning. The forecasting model is trained in two stages. In the first stage, the pretrained encoder and cross-attention decoder produce an initial mean forecast. In the second stage, this stage-1 model is frozen and two Patch-TST-style residual branches \cite{nie2023patchtst} are trained with the same architecture: one predicts a residual correction to the mean forecast, and the other predicts the uncertainty interval. These residual branches use patches of length 16 with stride 8 and are implemented as 6-layer, 8-head transformers with hidden size 512, feed-forward dimension 1,024, attention pooling, and dropout 0.1. They take the displacement context, the initial forecast, and metadata, imputation, and calendar features as additional inputs. 

For seismic step localization, the pretrained model is combined with a temporal DETR head. The encoded hidden states of width 768 are first projected to a decoder width of 256 and combined with a one-dimensional sinusoidal positional encoding. A transformer decoder with 4 layers, 8 attention heads, feed-forward dimension 1,024, and dropout 0.1 then processes $Q = 48$ learned queries. Each query carries both a learned content embedding and an initial reference point distributed across the 512-day window. Query locations are refined iteratively across decoder layers following Dynamic Anchor Box DETR (DAB-DETR \cite{Liu2022DABDETR}). In addition to the query-based branch, the model also predicts dense frame-level event scores along the encoded sequence and a window-level event probability. The dense frame branch gives local evidence for the query predictions and helps the model localize events more precisely within the window, while the binary branch suppresses query predictions in windows with no events. For detection, the pretrained model remains trainable and is optimized together with the detection head. This gives a detection model with 367 million parameters, of which 187 million are trainable. We use full finetuning for this task because seismic step localization requires stronger task-specific adaptation than forecasting. While forecasting is close to the pretraining objective of reconstructing masked parts, localization requires the encoder to associate preserved step-like patterns with sparse seismic-event labels and to localize their dates. 

\subsection{Evaluation Procedure and Metrics}
\label{subsec:eval_metrics}

We evaluate the model at three stages: self-supervised pretraining, displacement forecasting, and seismic step localization. 

For pretraining, we report validation performance under fixed masking conditions so that checkpoints are directly comparable even though the training curriculum changes over time. We use a validation probe at the final masking difficulty and a tail-mask probe in which the masked region is restricted to the end of the window. On these probes, we report the validation loss and cosine similarity between the predicted and target representations at masked positions. We also monitor codebook usage through hard-assignment perplexity and the fraction of used codevectors. 

In addition to these quantitative metrics, we perform an exploratory analysis of the learned codebook on 1,000 validation windows. For each window, we extract the discrete code assignments at every latent position and aggregate occurrences of each code across all windows. Each latent position corresponds to a local receptive-field patch in the input window rather than to a single exact day. We then characterize code occurrences by three interpretable properties: step sensitivity, trend sensitivity, and seasonal sensitivity. Step sensitivity is characterized by association with catalogued seismic events within the receptive field, together with the strength of the local step-like change. Trend sensitivity is characterized by the magnitude of the local displacement slope within the receptive field. Seasonal sensitivity is characterized by recurring code occurrences at annual or semi-annual periods.  

For forecasting, we evaluate 90-day displacement prediction using mean absolute error (MAE) and root mean squared error (RMSE). For the probabilistic second-stage model, we additionally report empirical coverage of the predictive intervals after calibration on the validation set. 

For seismic step localization, we evaluate both event detection and date localization. Because the labels in daily GNSS step catalogs can be misaligned by a few days relative to the visually apparent step, a predicted event is counted as correct if it falls within $\pm 2$ days of a ground-truth event. This is consistent with the error analysis of Crocetti et al. \cite{crocetti2021discontinuity}, who report that correctly classified earthquake dates often deviate by one to two days. We report three F1 metrics. The window F1 evaluates only whether a window contains at least one seismic event. The end-to-end localization F1 evaluates the full pipeline, so a prediction is counted as correct only if the model both predicts an event in the correct window and places the event date within the tolerance window. The localization F1 on target windows removes the window-level decision step and evaluates date localization only on windows that truly contain a ground-truth event. In addition, we report a conditional localization success rate on windows that contain both a ground-truth event and receive a positive event prediction.  

\section{Results}
\label{sec:experiments}

\subsection{Pretraining Analysis}

\begin{figure}[!t]
  \centering
  \includegraphics[width=\linewidth]{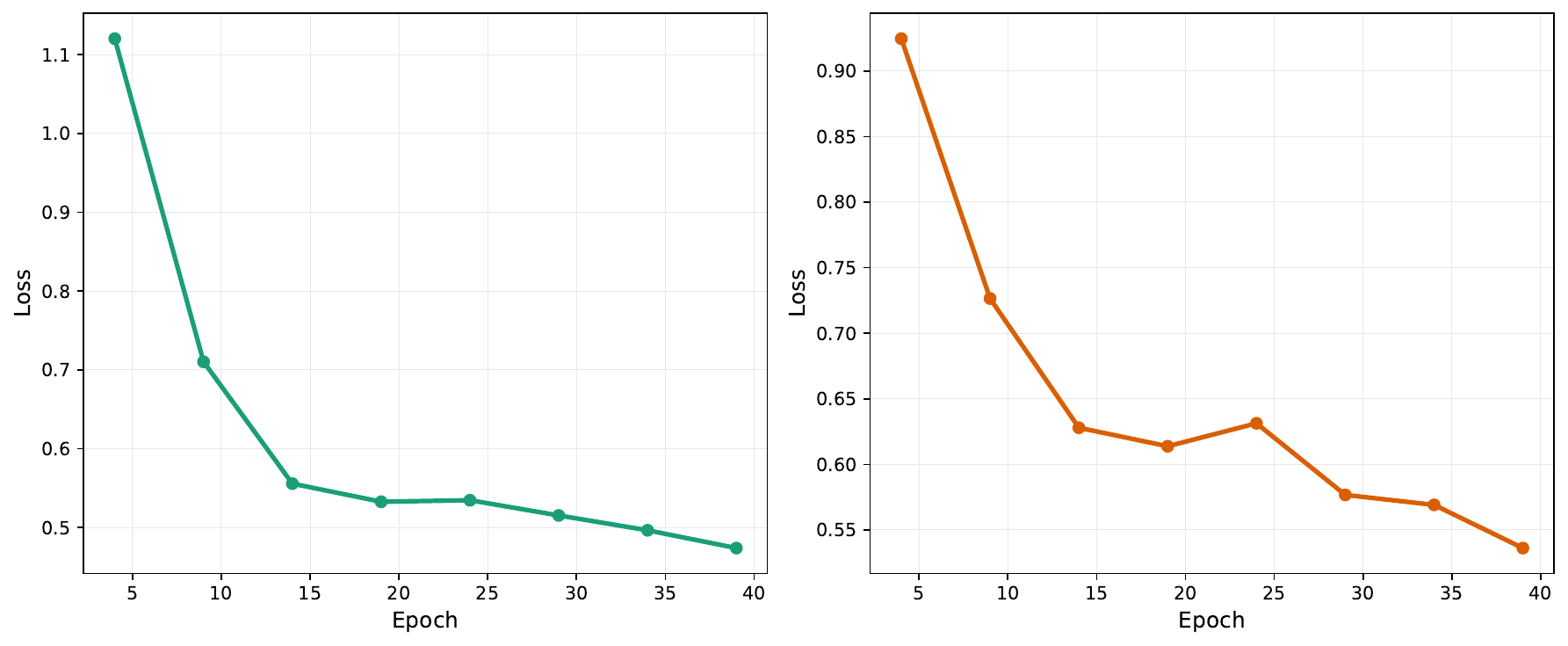}
  \caption{Validation under fixed masking conditions. Left: Fixed hard-mask probe loss over training. Right: Tail-mask probe loss over training. Both probes are evaluated with fixed masking difficulty at every epoch, so they are directly comparable across training.}
  \label{fig:pretrain_fixed_probe}
\end{figure}

To assess what the model learns during pretraining, we analyze the reconstruction quality, codebook usage, and the structure of the learned representations. The validation probes and metrics are defined in Section \ref{subsec:eval_metrics}. Figure~\ref{fig:pretrain_fixed_probe} shows that both validation probes improve steadily over training, indicating that the model learns to reconstruct masked regions increasingly well. The mean cosine similarity at masked positions increases from 0.849 to 0.902. On the more difficult tail-mask probe, cosine similarity rises from 0.713 to 0.803 for the last three masked positions. 

\begin{figure}[!t]
  \centering
  \includegraphics[width=\linewidth]{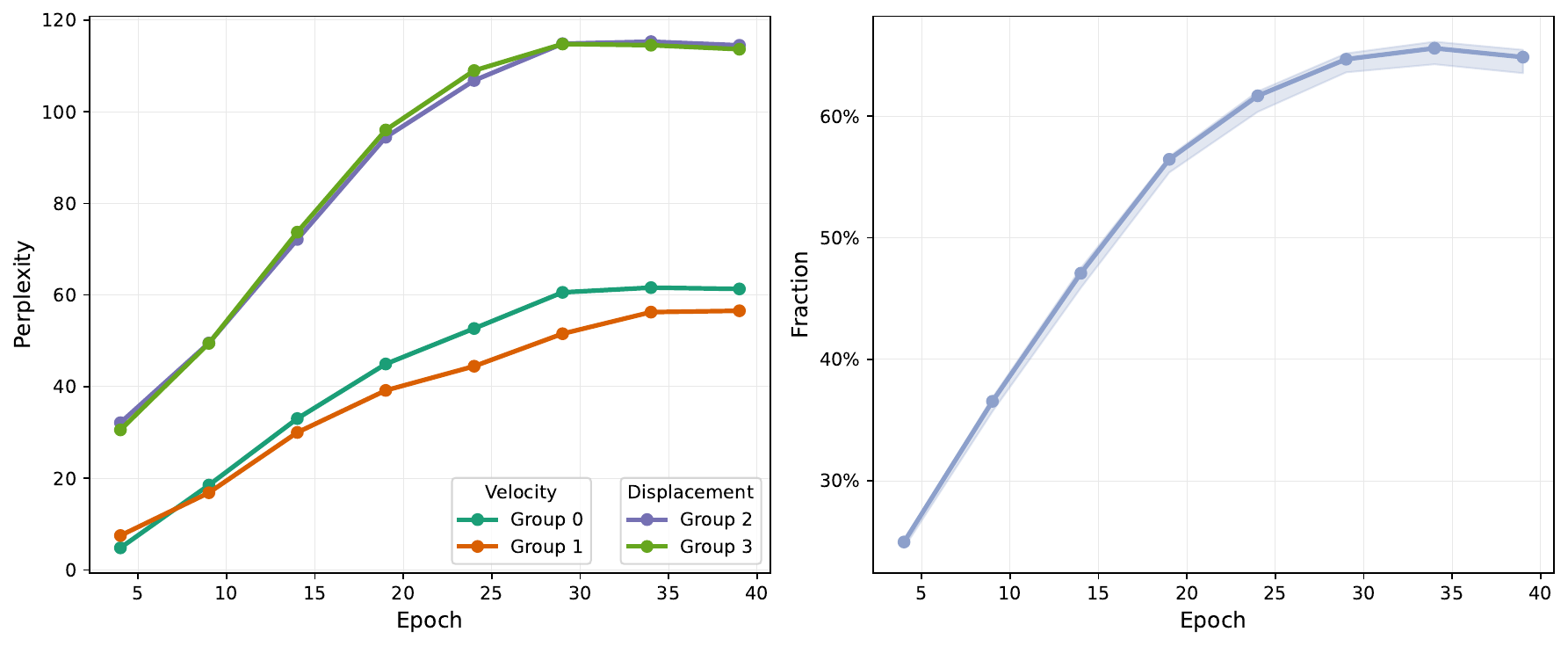}
  \caption{Quantizer usage over training. Left: Per-group hard-assignment perplexity across epochs. Right: Fraction of codevectors used on masked validation positions. The solid line shows the median across the validation interval, and the shaded band shows the interquartile range. Groups 0 and 1 correspond to the velocity stream, while groups 2 and 3 correspond to the displacement stream.}
  \label{fig:quantizer_usage}
\end{figure}

Figure \ref{fig:quantizer_usage} summarizes how codebook usage evolves over training. Hard-assignment perplexity increases steadily in all four groups (i.e., two velocity-related groups, and two displacement-related groups), showing that code usage becomes broader as training progresses. At the same time, the fraction of used codevectors also rises, which indicates that the model activates a larger part of the codebook over time. The velocity-related groups remain more selective than the displacement-related groups, while the displacement-related groups use a broader set of codevectors overall. This suggests that the two streams encode different kinds of information, with the displacement-related groups covering a more diverse range of target patterns. Within each stream, however, the two codebook groups do not show a clear separation by signal type. 

\begin{figure*}[!t]
  \centering
  \includegraphics[width=\linewidth]{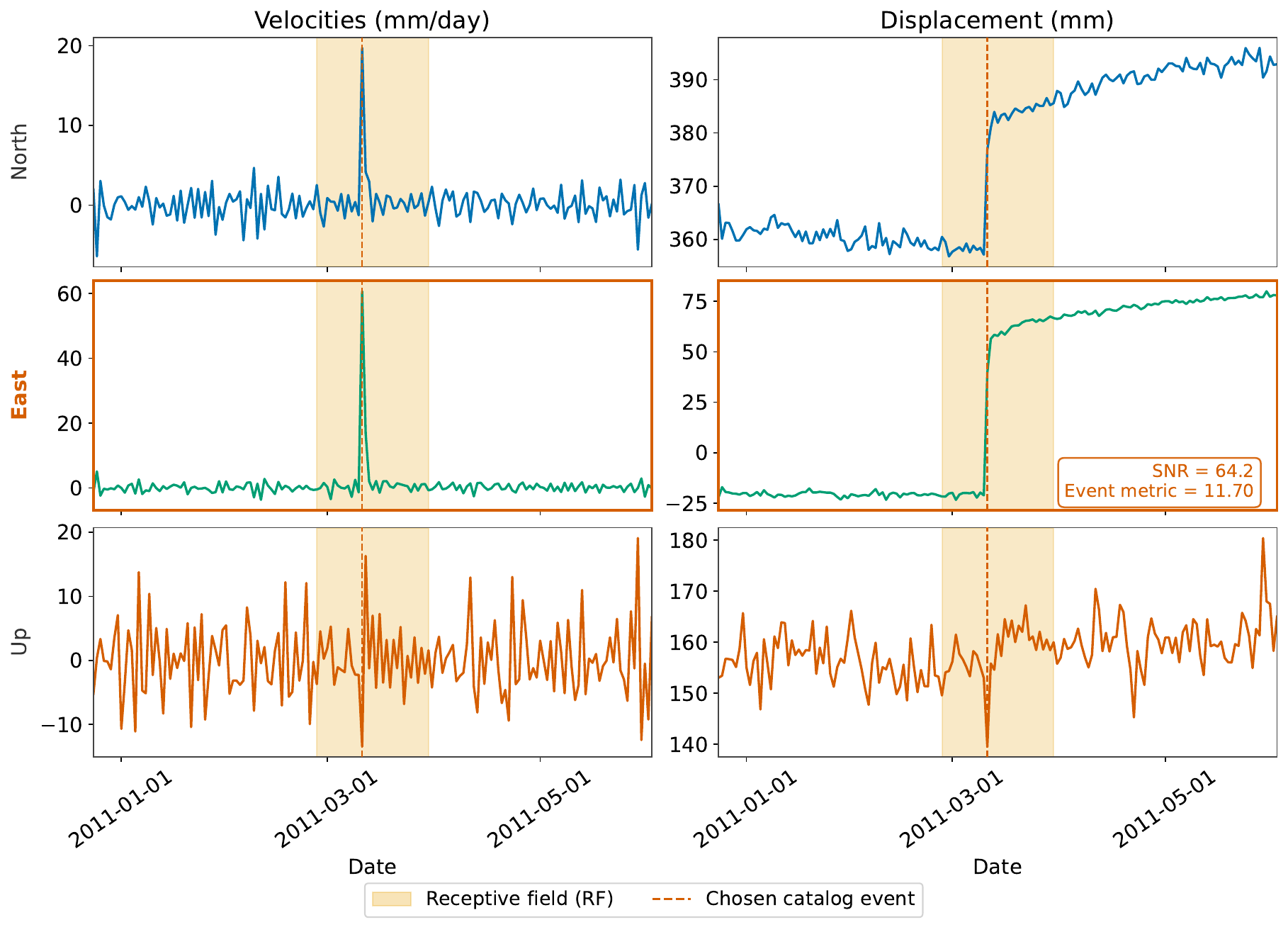}
    \caption{Example occurrence of the event-sensitive code g3-c221 at station J300. The shaded region shows the receptive field of the latent position assigned to this code, and the vertical line marks the matched catalog event. Although the code belongs to displacement group 3, the aligned velocities are also shown to illustrate the local change associated with the same time interval. The East component is highlighted because it has the strongest local step evidence, meaning the largest pre/post-event displacement offset relative to the local short-term variability. SNR denotes the local step signal-to-noise ratio, and Event metric the detectability z-score of the matched catalog event.}
  \label{fig:code_specialization_event}
\end{figure*}

The codebook analysis suggests that individual codevectors specialize in distinct signal types. To connect a codevector to the original time series, we map its assignments back to the corresponding receptive fields in the GNSS window. A code is considered event-sensitive if it is often assigned to positions where the receptive field overlaps with a catalog event. Figure \ref{fig:code_specialization_event} shows one representative example, code 221 from group 3. This code was selected because it appears near catalog-matched events more often than most other codes in the same group. The figure shows one such occurrence, where the code is assigned close to a seismic step. Similar analyses indicate that other codes are associated with trend-like or seasonal behavior. This suggests that the pretraining objective organizes the latent space around the main signal types present in GNSS data. 

\begin{figure}[!t]
  \centering
  \includegraphics[width=\linewidth]{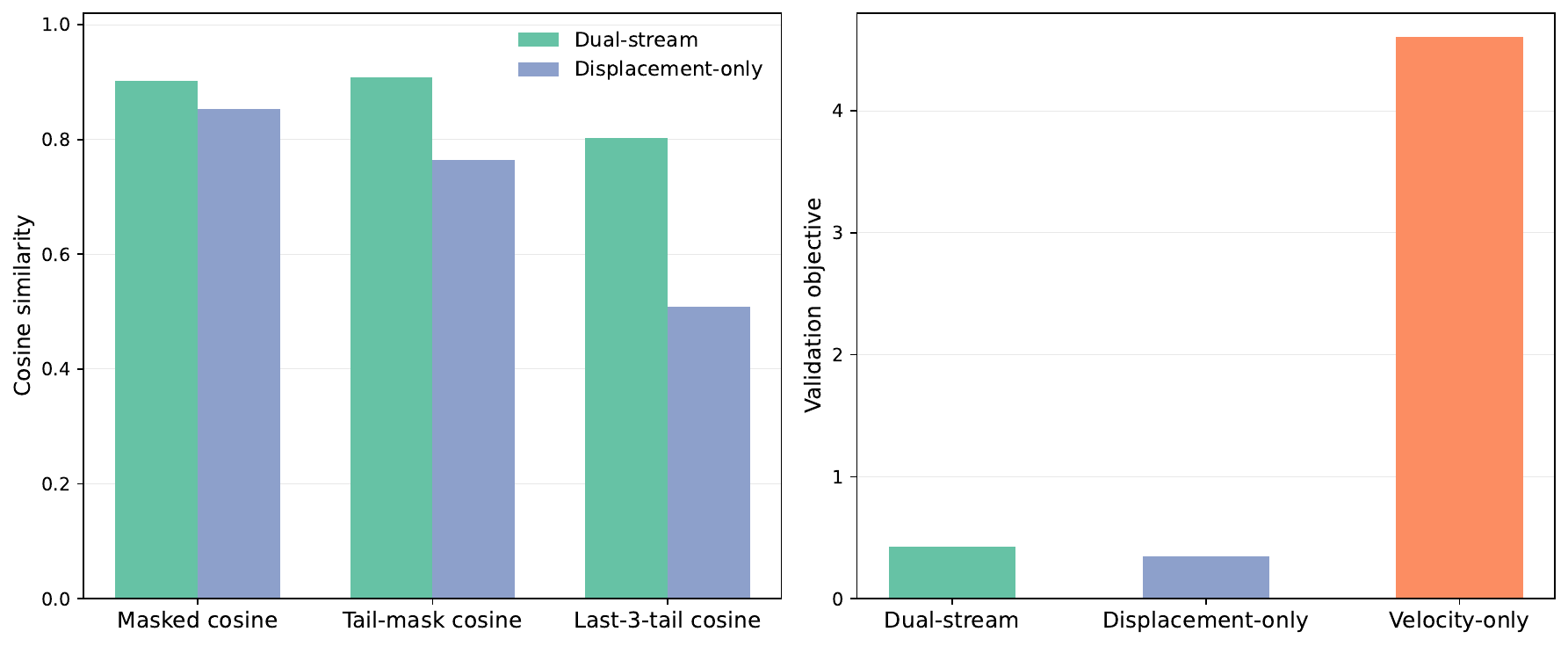}
  \caption{Comparison of the dual-stream model with the single-stream ablations. Left: Cosine-based probe metrics, where 1 indicates perfect alignment between the predicted and target representations. Right: Overall validation objective, where lower values indicate better optimization.}
  \label{fig:pretrain_ablations}
\end{figure}

We also train two ablation models using only the velocity or only the displacement stream. Figure \ref{fig:pretrain_ablations} compares these models with the full dual-stream model. The velocity-only model is dominated by codebook collapse and fails to optimize the objective successfully, so its cosine similarity values are much less informative. The displacement-only model remains much stronger, but it performs worse than the dual-stream model on all cosine-based probe metrics, with the largest gap on the tail-mask probes. The slightly lower overall validation objective of the displacement-only model should be interpreted cautiously, because this objective combines the main contrastive term with additional regularization terms and does not correspond directly to probe cosine similarity quality. In particular, the displacement-only objective excludes the velocity-stream targets and their associated code-usage regularization, which can make the total objective slightly lower without implying better representations. Overall, this confirms that both streams contribute and that the velocity stream appears to be especially important for extrapolation. 

\subsection{Displacement Forecasting}

We compare two supervised baselines and two finetuned variants of our pretrained model. The baselines are the Calibrated Base Forecast, which combines ridge regression and Fourier extrapolation, and the Patch-TST Residual Baseline \cite{nie2023patchtst}, which adds a supervised residual corrector on top of that calibrated base forecast. The two variants of our model are \gnssfm{} Stage 1, which combines the pretrained encoder with a cross-attention decoder, and \gnssfm{} Stage 2, which adds a Patch-TST residual branch.  

Table \ref{tab:forecasting} shows the main forecasting results. The two supervised baselines achieve lower MAE than \gnssfm{}, whereas \gnssfm{} Stage 1 achieves the lowest RMSE on both validation and test set. The baselines are designed to extrapolate recent trend and seasonal harmonics directly, which makes them very strong on the many ordinary windows where the future is a smooth continuation of the context. \gnssfm{}, in contrast, predicts through a downsampled pretrained representation and is selected with an RMSE-oriented validation criterion, which favors reducing rare large forecast failures over minimizing the typical absolute error measured by MAE. The Up component produces the largest MAE across all models, while North and East contribute more to RMSE through rare large errors from seismic events. Stage 2 changes the validation metrics only slightly: the RMSE decreases by 0.13\%, while the MAE increases by 1.60\%. Since this does not represent a meaningful improvement of the deterministic mean forecast, we retain Stage 1 as the final point-forecasting model and interpret Stage 2 as the probabilistic extension for uncertainty estimation. After recalibration on the validation set, the empirical coverages of the 68\%, 90\%, and 95\% predictive intervals on the test set are 67.2\%, 89.6\%, and 94.8\%, respectively. 

\begin{figure}[!t]
  \centering
  \includegraphics[width=\linewidth]{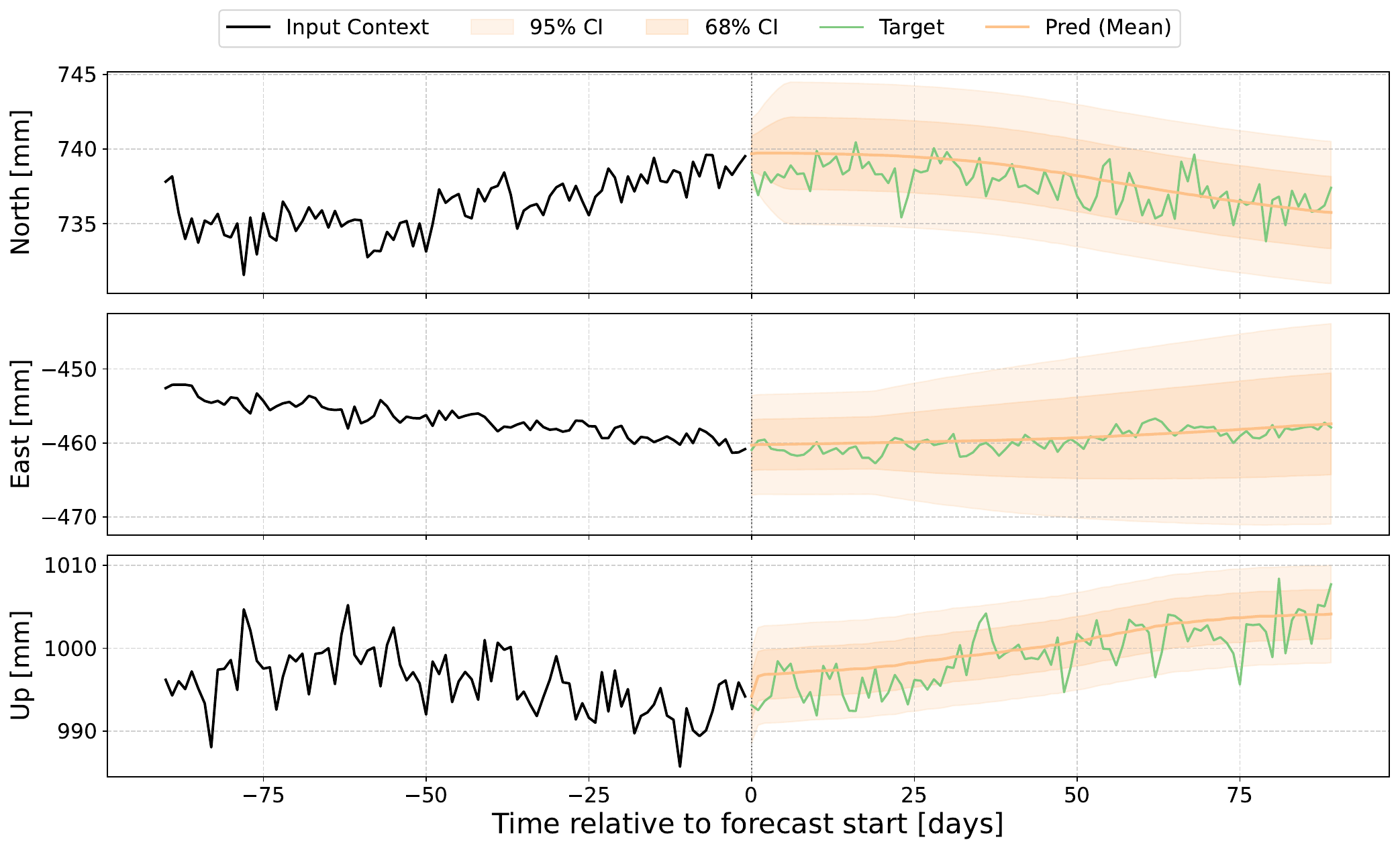}
  \caption{Representative probabilistic forecast from \gnssfm{} Stage 2 on the test set for station NOSE. The figure shows, for the North, East, and Up displacement components, part of the input context used by the model (black), the ground-truth future displacement over the 90-day forecast horizon (green), the predicted mean trajectory (orange), and the predictive confidence intervals (shaded areas).}
  \label{fig:forecast_example_gnssfm}
\end{figure}

The difference between validation and test is much larger for the two supervised baselines than for \gnssfm{}. In particular, RMSE increases strongly from validation to test for both baselines, while the pretrained model remains much more stable across the two splits. Figure \ref{fig:forecast_example_gnssfm} shows a representative probabilistic forecast from \gnssfm{} Stage 2 on the test set. The predicted mean captures the low-frequency evolution of the displacement time series, while remaining smoother than the daily observations. This behavior is expected for daily GNSS forecasting, where small day-to-day fluctuations are often dominated by measurement noise and are difficult to predict from the context alone. The predictive intervals therefore represent the remaining short-term variability around the smoother mean trajectory. 

\begin{table}[!t]
  \centering
  \caption{Displacement forecasting results in millimetres for the different models on the validation and test set. \gnssfm{} Stage 1 is retained as the final point-forecasting model, while Stage 2 is evaluated as the probabilistic extension.}
  \label{tab:forecasting}
  \setlength{\tabcolsep}{5pt}
  \begin{tabular}{|l|l|r|r|}
    \hline
    Split & Model & MAE & RMSE \\
    \hline
    Val  & Base forecast       & \textbf{3.67} & 22.84 \\
    Val  & Patch-TST baseline  & 3.88 & 19.71 \\
    Val  & \gnssfm{} Stage 1   & 5.61 &  7.91 \\
    Val  & \gnssfm{} Stage 2   & 5.70 &  \textbf{7.90} \\
    \hline
    Test & Base forecast       & \textbf{4.04} & 77.05 \\
    Test & Patch-TST baseline  & 4.17 & 58.77 \\
    Test & \gnssfm{} Stage 1   & 4.87 &  \textbf{6.78} \\
    Test & \gnssfm{} Stage 2   & 5.10 &  6.98 \\
    \hline
  \end{tabular}
\end{table}

\subsection{Seismic Step Localization}

We evaluate seismic step localization on 512-day windows. The task combines a window-level event decision with event date localization, and the evaluation metrics are defined in Section \ref{subsec:eval_metrics}. We compare \gnssfm{} with a hybrid baseline that uses multi-scale Haar step scores \cite{mallat2008wavelet} and derivative-energy features for candidate proposals, XGBoost \cite{chen2016xgboost} for ranking, and a small temporal convolutional network for refinement. 

\begin{figure}[!t]
  \centering
  \includegraphics[width=\linewidth]{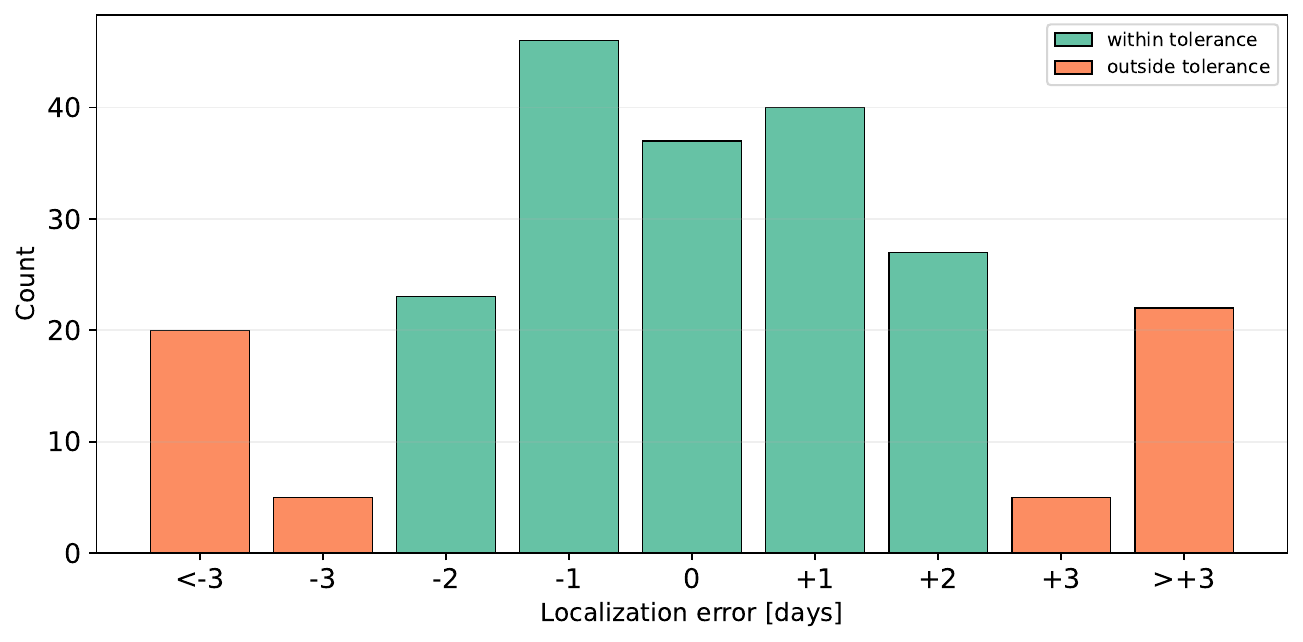}
  \caption{Histogram of localization errors on windows that both contain a ground-truth event and receive a positive event prediction. Green bars mark errors within the $\pm 2$-day tolerance used for the localization metrics, while orange bars are outside this tolerance. The concentration near zero shows that, once the model predicts an event in a true event window, the predicted date is often close to the ground-truth date.}
  \label{fig:localization_error_hist}
\end{figure}

\begin{figure}[!t]
  \centering
  \includegraphics[width=\linewidth]{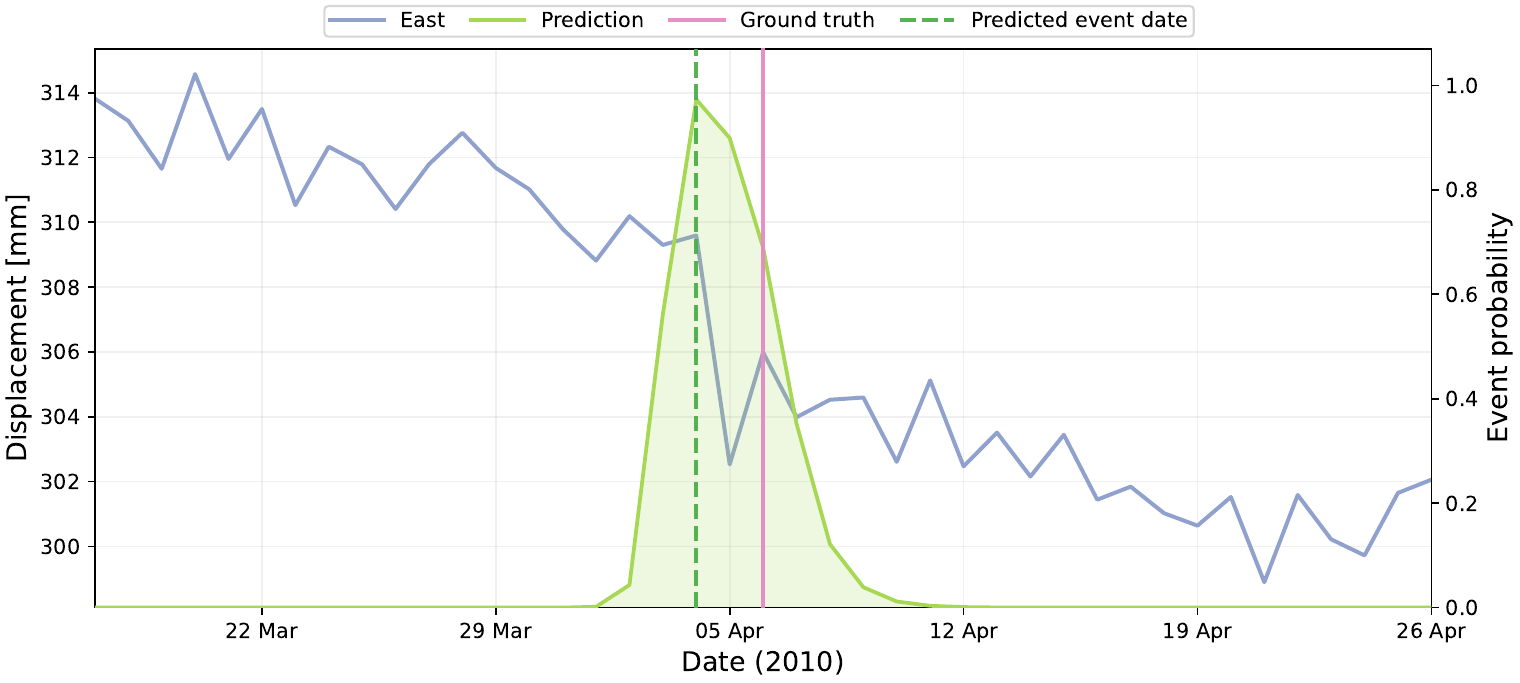}
  \caption{Representative seismic step localization for one example in the test set. The figure shows the displacement component with the highest signal-to-noise ratio and the predicted event probability over time. For this window of station P482, the binary detection branch predicts an 82.4\% probability of a seismic event. The vertical lines mark the predicted event date and the ground-truth seismic step date. The predicted event probability peaks two days before the catalog event date, but aligns with the onset of the visible displacement change, illustrating the $\pm 2$-day matching tolerance used for evaluation.}
  \label{fig:detection_example_window}
\end{figure}

Table \ref{tab:detection} summarizes the results on the test set. The two models are similar at the window-level event decision, but differ much more strongly at the localization level. The main improvement of \gnssfm{} therefore does not come from deciding whether a window contains an event at all, but from turning that decision into an accurate event date. For \gnssfm{}, the conditional localization success rate is 0.769 on windows that both contain a ground-truth event and receive a positive event prediction. This shows that once the model predicts an event in a true event window, it often places the event date close to the ground-truth date. Figure \ref{fig:localization_error_hist} complements this by showing the distribution of localization errors around the ground-truth date. For reference, when the model is optimized specifically for either window classification or target-only localization, it reaches an F1 score of 0.85 and 0.69, respectively. A representative example window is shown in Figure \ref{fig:detection_example_window}, which illustrates the predicted event probability and the predicted event date relative to the ground-truth seismic step. 

\begin{table}[!t]
  \centering
  \caption{Seismic step localization results on the test set. The three F1 metrics are defined in Section \ref{subsec:eval_metrics}.}
  \label{tab:detection}
  \setlength{\tabcolsep}{4pt}
  \begin{tabular}{|l|r|r|r|}
    \hline
    Model & Window F1 & End-to-end F1 & Localization F1 \\
    \hline
    Baseline & 0.574 & 0.035 & 0.175 \\
    \gnssfm{}      & \textbf{0.590} & \textbf{0.429} & \textbf{0.515} \\
    \hline
  \end{tabular}
\end{table}

\section{Discussion}
\label{sec:discussion}

\subsection{Representation Learning in the Pretrained Encoder}

The pretraining results suggest that the model learns more than a simple interpolation of missing values. The appearance of separate codevectors for seismic offsets, trends, and seasonal cycles reflects the classical decomposition used in geodetic time series analysis, where the signal is explicitly separated into trend, periodic, offset, and noise components \cite{williams_effect_2003, dong_seasonal_2002}. The pretrained encoder produces a similar structure implicitly, which suggests that the main signal types in GNSS data repeat often enough to be captured by a discrete latent vocabulary. 

The ablation results also clarify what each input stream contributes. Displacement carries most of the stable structure needed for masked reconstruction, while the velocity stream is especially important for the more difficult segments. This is physically plausible: drift, seasonal loading, and step offsets are more directly visible in the displacements, whereas the velocity stream provides short-term dynamics that are otherwise suppressed by the convolutional downsampling. 

The main limitation of the current pretraining approach is that, although the model conditions on station metadata, it processes each station window independently. The metadata provides static spatial information, but it does not provide the model with simultaneous observations from neighboring stations or with time-varying external geophysical drivers. This matters because shared signals such as atmospheric loading affect neighboring stations simultaneously and are a well-known challenge in GNSS network analysis \cite{wdowinski_southern_1997, dong2006kle}. Including spatial context, for example through a graph-based encoder or cross-station attention, would likely help, especially for the vertical component where loading effects and common-mode errors are often strongest. 

\subsection{Displacement Forecasting}

The MAE-RMSE trade-off in Table \ref{tab:forecasting} shows that the pretrained model changes the pattern of errors rather than improving all windows. This is because convolutional downsampling in the pretrained encoder suppresses high-frequency variability, producing smoother forecasts that avoid large errors, but also miss some short-term signals in easy windows. This trade-off is beneficial overall because a small number of very difficult windows, caused by seismic steps or aggressive local drifts, dominate the squared error far more than they affect the absolute error. 

\begin{figure}[!t]
  \centering
  \includegraphics[width=\linewidth]{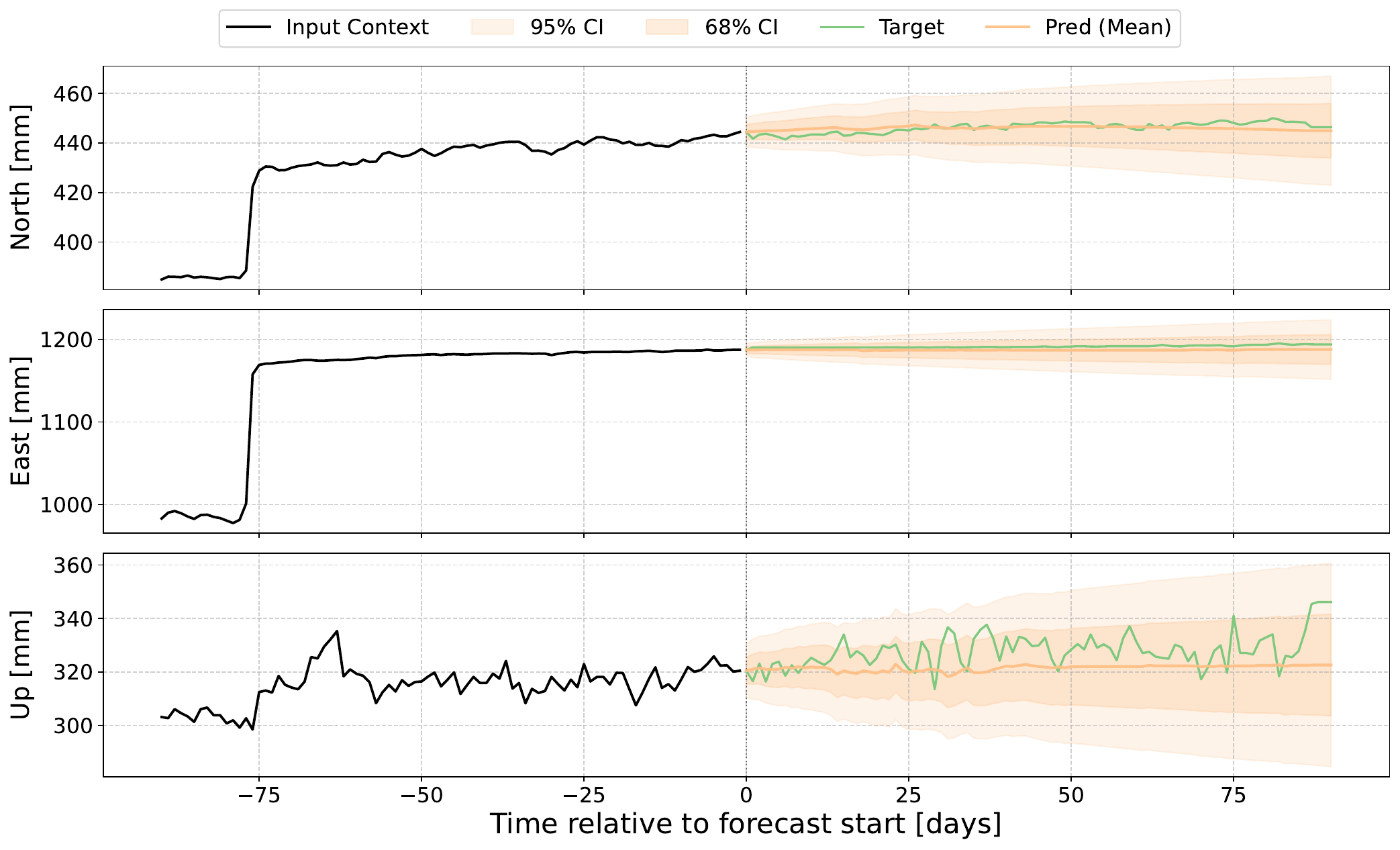}
  \caption{Forecast example for station J263 illustrating the behavior of \gnssfm{} Stage 2 after a seismic step in the input context.}
  \label{fig:forecast_step_context}
\end{figure}

The much larger gap between validation and test RMSE for the supervised baselines suggests that the test split contains a higher fraction of such difficult forecast windows. This interpretation is supported by the fact that baseline MAE changes only slightly between validation and test, whereas baseline RMSE increases strongly. This indicates that the main difference comes from rare, high-error cases rather than from general overfitting across typical windows. In particular, the baselines appear to be more sensitive when abrupt changes distort the recent trend near the context boundary or when a short local drift is extrapolated too aggressively. By contrast, the pretrained model remains much more stable across the two splits. Figure \ref{fig:forecast_step_context} illustrates this behavior for a difficult test example with a seismic step in the context window. The forecast remains stable instead of extrapolating the abrupt offset into the prediction horizon, while still tracking the broad North and East trajectories. The Up component is noisier, and the model represents much of this short-term variability through wider predictive confidence intervals rather than the mean trajectory. Since Stage 2 mainly adds probabilistic information without substantially changing the mean forecast, a natural extension would be to decouple the low-frequency point forecast from the high-frequency residual variability more explicitly, for example by preserving the Stage 1 mean prediction and training a separately calibrated uncertainty branch. 

All forecasting approaches share the same hardest failure mode: seismic steps that occur within the forecast horizon. These are impossible to predict from the previous displacement time series alone. As a result, forecast errors around such events reflect a limitation of the available input information rather than only a weakness of the forecasting architecture. 

\subsection{Seismic Step Localization}

The large improvement in end-to-end localization F1 over the baseline reflects a key difference in how the two systems work. The baseline depends on a proposal stage that must identify the correct candidate event date before any learning-based ranking can take place. When this stage fails, as it does for many events, the downstream ranker has nothing to work with. The pretrained model avoids this bottleneck by operating directly on the learned sequence representation, where event-sensitive codevectors already encode step information. 

The final joint model reaches a test window-level F1 score of 0.590 and a test localization F1 on target windows of 0.515. The corresponding task-specific reference results are higher, with a test window-level F1 score of 0.85 for the window-event-only model and a test localization F1 on target windows of 0.69 for the localization-focused model. This gap suggests that the two objectives create partially conflicting demands during training. Better loss balancing, staged training, or architectural separation of the two heads are the most promising directions for closing this gap. 

\begin{figure}[!t]
  \centering
  \includegraphics[width=0.9\linewidth]{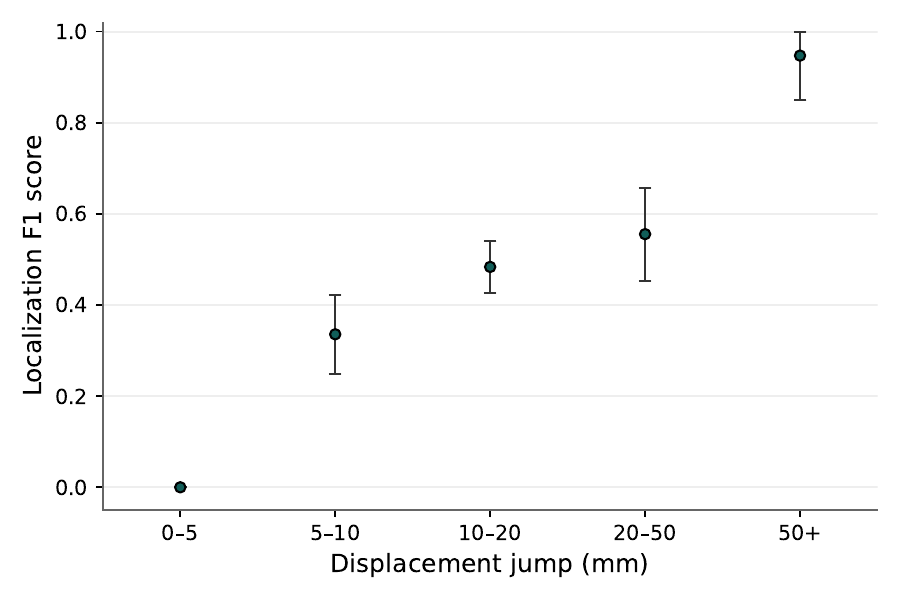}
      \caption{Localization performance on true-event windows in the test set, stratified by event size. Event size is measured as the displacement jump at the event date. The points show localization F1 score in each size bin, and the error bars show 95\% percentile bootstrap confidence intervals obtained by resampling true-event windows within each bin. The size bins are defined as $[0,5)$, $[5,10)$, $[10,20)$, $[20,50)$, and $\geq 50$ mm, with 12, 114, 301, 91, and 20 true-event windows in each bin, respectively. Localization performance increases clearly with event size, showing that weak observed offsets near the detectability threshold are much harder to recover than larger seismic steps.}
  \label{fig:detection_size_stratified_discussion}
\end{figure}

A further limitation specific to the detection task is label quality. The ground-truth event catalog is based on a simple distance and magnitude equation that produces many candidate events that are not actually detectable in the displacement data. While we filter these using a signal-to-noise threshold, the remaining labels are still imperfect: some retained events may still not produce a visible step, some are shifted by a few days, and some real steps may be absent from the catalog entirely. This likely causes the evaluation metrics to underestimate true model performance. At the same time, an analysis using the observed displacement jump on the event date shows a clear dependence on event strength: localization F1 rises steadily with displacement jump size, from 0.34 for 5-10 mm offsets to 0.56 for 20-50 mm, and reaches 0.95 above 50 mm (Figure \ref{fig:detection_size_stratified_discussion}). Consistent with this trend, the miss rate is higher for the smallest displacement jumps and declines strongly for larger events, whereas conditional localization precision remains comparatively high once the model emits a detection. The main limitation of the current detector is recall for weak seismic offsets near the detectability threshold, rather than date placement once an event has been detected. 

\subsection{Limitations and Future Directions}

The most important limitation of \gnssfm{} is that pretraining uses a single data source. Adding more GNSS networks, which rely on different processing pipelines, or additional geodetic data could improve generalization, especially in regions with fewer stations. The daily temporal resolution also limits the model to applications that do not require high-frequency data. 

Beyond the two downstream tasks presented here, the pretrained encoder is a natural candidate for GNSS data gap imputation, denoising, anomaly detection, and station quality evaluation, thus, illustrating the broad potential of \gnssfm{}. In principle, it is suitable for all applications where labeled data are rare and a pretrained representation could reduce the supervision needed. 

\section{Conclusion}
\label{sec:conclusion}

We present \gnssfm{}, a self-supervised foundation model for daily GNSS displacement time series. We pretrain a dual-stream transformer encoder using masked latent prediction with vector-quantized targets, adapted from wav2vec 2.0 with several modifications for GNSS data. Pretrained on data from over 17,000 globally distributed GNSS stations, the model learns meaningful representations of GNSS signals, as suggested by codebook entries that specialize in seismic offsets, trends, and seasonal patterns. 

On two downstream tasks with very different task formulations, the pretrained encoder outperforms supervised baselines in both cases. In displacement forecasting, the finetuned model achieves a test RMSE of 6.78 mm compared to 58.77 mm for the supervised Patch-TST baseline. In seismic step localization, the end-to-end localization F1 improves from 0.035 (baseline) to 0.429, with a median absolute localization error of two days on correctly identified event windows. 

Together, these results demonstrate that a single pretrained GNSS backbone can support multiple downstream tasks, reduces the dependence on labeled data, and generalizes well across stations it has not seen during pretraining or finetuning, thus showing great potential for future applications. 

\section*{Acknowledgements}
The contributions of Fanny Lehmann to this work were primarily supported by the ETH AI Center fellowship. The contributions of Leonardo Trentini were supported by the Swiss National Science Foundation (SNSF) under grant number 225851.

\appendices
\section{Preprocessing Details}
\label{app:preprocessing}

This appendix summarizes the detailed preprocessing steps that are outlined in Section \ref{sec:data}. The goal is to transform daily GNSS displacement time series into fixed-length training windows while preserving the main deformation structure and keeping track of the data quality. Unless stated otherwise, all preprocessing steps are applied separately to the north, east, and up components. 

\subsection{Station filtering}

Before the final station selection, poorly observed edges are trimmed to remove fragmented boundary regions. The final retained time span is the intersection across north, east, and up component. Stations are then filtered using simple criteria based on observation span, overall coverage, and long internal gaps. One station with unusually large seasonal amplitudes (LTHW) is removed separately. 

\subsection{Outlier detection and event handling}

Outliers are treated as isolated spikes that are inconsistent with the local behavior of the displacement series and that do not persist as step-like offsets. Detected outliers are masked so that later gap filling treats them in the same way as missing samples. 

The outlier detector is an iterative procedure based on a Hampel filter, combined with additional tests on local jumps and large deviations from a detrended series \cite{pearson2016generalized,davies1993identification}. Candidate outliers that occur close together are grouped, and the displacement before and after each group is compared. A group is kept only if it is consistent with a real offset rather than an isolated spike. This helps to preserve seismic or other step-like changes while still removing outliers. The comparison between the two surrounding windows is based on a two-sided Mann-Whitney test \cite{mann1947test}. 

For each cataloged event, a local detectability score is computed from the east/north/up time series in a short window around the catalog date. Equipment changes with a detectability score above a threshold are masked before gap filling, because otherwise, after differencing they would appear as isolated spikes. After preprocessing, the event refinement pass is repeated on the cleaned time series to update detectability scores. Only seismic events are used as positive targets in the downstream detection task. 

\subsection{Velocity stream, normalization, and gap filling}

After outlier removal, the cleaned displacement series is converted into first differences. This representation reduces non-stationarity and makes local changes more explicit. The velocity stream is normalized before gap filling. The series is then transformed by an inverse hyperbolic sine transformation to compress large values without clipping. 

Missing samples in the velocity stream, including removed outliers and masked equipment changes, are filled in two ways. Short gaps are filled by linear interpolation. Longer gaps are filled by a resampling-based synthesis step that uses nearby valid data to maintain local variability \cite{kunsch1989jackknife,lall1996nearest}. Each day also receives a reliability label indicating whether it is original, interpolated, synthesized, or padded. 

\subsection{Displacement reconstruction and dual-stream output}

After filling gaps in the velocity stream, the displacement stream is rebuilt by summing the daily increments. In long gaps, the reconstructed values are adjusted so that they still match the observed displacement on both sides of the gap. 

The final model input contains two aligned streams. The velocity stream emphasizes local and fast-changing behavior, while the displacement stream preserves the absolute displacement level and longer-range dependencies. The two streams are normalized separately because they differ in scale and distribution. 

\subsection{Window construction and reliability masks}

All experiments use fixed-length windows of 512 days. For pretraining and forecasting, windows are non-overlapping. For seismic step localization, a limited overlap is used so that events near window boundaries are less likely to be missed. 

To avoid windows dominated by imputed values, only windows with sufficient original data are kept. Short series may still produce one padded window, with the padded region excluded later through masking. For forecasting, each window is split into a context period and a 90-day prediction horizon, and the horizon must also satisfy a minimum original-data requirement. 

Reliability labels are converted into the masks and weights used during training. These separate original from padded samples and reduce the influence of imputed values in the loss. 

\subsection{Detection-specific augmentation and batch balancing}

Detection training includes two additional steps because seismic events are rare relative to the total number of windows. First, windows that contain seismic events are augmented by small random time shifts. Second, the detection uses balanced batches so that positive windows are seen regularly during training despite the strong class imbalance.

\section{Pretraining Objective Details}
\label{app:pretraining_details}

This appendix gives the definitions of the terms in Equation \ref{eq:full_pretrain_loss}. The pretraining objective combines the masked contrastive term with a diversity term, a small auxiliary reconstruction term, and several regularization terms that reduce codebook collapse. 

The main term is the contrastive loss. For each masked feature step $t \in \mathcal{M}$, the model compares the projected transformer output $c_t$ with the corresponding quantized target $q_t$ and with $K=50$ negative targets $q_{t,j}^{-}$ sampled from other feature steps. The negatives are taken both from the current batch and from the memory queue. The contrastive term is weighted by the reliability of the masked step, so masked targets based on original observations contribute more than targets based on imputed values. 

The diversity term $\mathcal{L}_{\mathrm{div}}$ encourages broad use of the codebook instead of collapse to a small number of codevectors \cite{baevski_wav2vec_2020}. The usage-entropy term $\mathcal{H}_{\mathrm{usage}}$ acts in a similar way and further encourages the average assignment distribution in each group to remain spread across many codevectors. In contrast, the token-entropy term $\mathcal{H}_{\mathrm{tok}}$ acts at the level of individual feature steps and discourages assignments from becoming too sharp too early. 

The auxiliary reconstruction term $\mathcal{L}_{\mathrm{aux}}$ is a small reliability-weighted loss applied only at masked positions. It consists of three L1 terms: one compares the predicted velocity with the target velocity, one compares the predicted displacement with the target displacement, and one compares the cumulative sum of the predicted velocity with the predicted displacement. This term helps stabilize training and keeps the latent representation tied to the underlying signal structure. 

The orthogonality term $\mathcal{L}_{\mathrm{orth}}$ acts on the raw codevectors and encourages them to point in different directions within each group. The separation term $\mathcal{L}_{\mathrm{sep}}$ acts on the projected codevectors after the trainable projection layer and keeps them separated in the contrastive space. This helps prevent collapse at both stages. 

The total-variation term $\mathcal{L}_{\mathrm{tv}}$ reduces large changes in the assignment probabilities from one feature step to the next. This makes the assignments smoother over time. 

Two additional regularizers are used for the velocity groups. The term $\mathcal{L}_{\mathrm{vh}}$ encourages the model to keep several code options active, and $\mathcal{L}_{\mathrm{vkl}}$ keeps the soft assignment distribution close to uniform so that the model does not rely too much on a small subset of codes. In addition, the term $\mathcal{L}_{\mathrm{lat}}$ penalizes low cosine similarity between the predictions and the corresponding targets. 

Not all coefficients are constant during training. The coefficients $\lambda_{\mathrm{aux}}(s)$, $\lambda_{\mathrm{vh}}(s)$, $\lambda_{\mathrm{vkl}}(s)$, and $\lambda_{\mathrm{lat}}(s)$ depend on the training step $s$ and are introduced gradually. This keeps the main contrastive objective dominant early in training and adds the auxiliary regularization terms later. 

In the final pretraining setup, the active coefficients are $\lambda_{\mathrm{div}}=1.0$, $\lambda_{\mathrm{aux}}(s)\leq 0.01$, $\lambda_{\mathrm{orth}}=0.05$, $\lambda_{\mathrm{tv}}=0.01$, $\lambda_{\mathrm{vh}}(s)\leq 0.04$, $\lambda_{\mathrm{vkl}}(s)\leq 0.05$, $\lambda_{\mathrm{lat}}(s)\leq 0.02$, $\lambda_{\mathrm{sep}}=0.05$, $\lambda_{\mathrm{usage}}=0.06$, and $\lambda_{\mathrm{tok}}=0.02$. 

\bibliographystyle{IEEEtran}
\bibliography{references}

\end{document}